\documentclass[aps,prl,twocolumn,superscriptaddress,amsmath,amssymb,longbibliography]{revtex4-1}
\usepackage[english]{babel}
\usepackage{graphicx,bbm}
\usepackage{mathtools}
\usepackage{times}
\usepackage[utf8]{inputenc}
\usepackage[colorlinks,linkcolor=blue,citecolor=blue,urlcolor=blue]{hyperref}
\usepackage{multirow}
\newcommand{\ms}[1]{\textcolor{black}{#1}}
\newcommand{\jm}[1]{\textcolor{black}{#1}}

\usepackage[normalem]{ulem}
\begin{document} 

\title{Hund bands in spectra of multiorbital systems}

\author{M. {\'S}roda}
\affiliation{Institute of Theoretical Physics, Faculty of Fundamental Problems of Technology, Wroc\l aw University of Science and Technology, 50-370 Wroc\l aw, Poland}

\author{J. Mravlje}
\affiliation{Jo\v{z}ef Stefan Institute, SI-1000 Ljubljana, Slovenia}

\author{G. Alvarez}
\affiliation{Computational Sciences and Engineering Division, Oak Ridge National Laboratory, Oak Ridge, Tennessee 37831, USA}

\author{E. Dagotto}
\affiliation{Department of Physics and Astronomy, University of Tennessee, Knoxville, Tennessee 37996, USA}
\affiliation{Materials Science and Technology Division, Oak Ridge National Laboratory, Oak Ridge, Tennessee 37831, USA}

\author{J. Herbrych}
\affiliation{Institute of Theoretical Physics, Faculty of Fundamental Problems of Technology, Wroc\l aw University of Science and Technology, 50-370 Wroc\l aw, Poland}

\date{\today\\[1.5em] }
%------------------------------------------------------------------------------------
\begin{abstract}
Spectroscopy experiments are routinely used to characterize the behavior of strongly correlated systems. An in-depth understanding of the different spectral features is thus essential. Here, we show that the spectrum of the multiorbital Hubbard model exhibits unique Hund \ms{bands} that occur at energies given only by the Hund coupling $J_\mathrm{H}$, as distinct from the Hubbard satellites following the interaction $U$. We focus on experimentally relevant single-particle and optical spectra that we calculate for a model related to iron chalcogenide ladders. The calculations are performed via the density-matrix renormalization group and Lanczos methods. The generality of the implications is verified by considering a generic multiorbital model within dynamical mean-field theory.\vspace{3em}
\end{abstract}
%------------------------------------------------------------------------------------
\vphantom{}
\maketitle

%------------------------------------------------------------------------------------
{\it Introduction.} 
Strongly correlated systems are at the heart of modern condensed matter physics. The celebrated single-band Hubbard model, describing (doped) Mott insulators, is still extensively studied in the context of Cu-based high-temperature superconductivity~\cite{Scalapino2012,Dagotto1994,Shane2022}. Equally exciting case is that of iron-based superconductors where the presence of several active orbitals leads to novel effects beyond the ``standard'' Mott physics~\cite{Yin2011,Georges2013,Yi2017}. A nontrivial example is the orbital-selective Mott phase (OSMP)~\cite{koga04,medici05,Georges2013,Rincon2014,Herbrych2019}, where Mott-localized and itinerant electrons coexist.

A key probe of electronic excitations is the single-particle spectral function $A(k,\omega)$, characterizing the excitations' dispersion. It is experimentally accessible by angle-resolved photoemission spectroscopy (ARPES)~\cite{Damascelli2003,Wang2020}. To understand the origin of different spectral features, it is convenient to consider idealized models that can be studied theoretically and monitor how the signatures of correlations (e.g., the Hubbard bands) evolve with increasing Coulomb interaction $U$. This is especially true for quantum systems of reduced dimensionality, for which quasiexact numerical methods~\cite{Benthien2004,Feiguin2011}, or even closed analytical solutions~\cite{Essler2010}, provide unbiased information on the elementary excitations. However, even in reduced dimensionality obtaining accurate results for the multiorbital Hubbard model remains challenging. The difficulty lies in the exceptionally large Hilbert space. Because of that, the spectral functions are often calculated using the dynamical mean-field theory (DMFT)~\cite{georges96,Jakobi2013,Roekeghem2016,tamai19}.
\jm{This approach, that strictly applies at large dimensionality, avoids the finite-size limitation, but often relies on solvers in Matsubara frequencies and hence the resulting spectral functions are blurred due to analytical continuation (see Ref.~\cite{Bauernfeind2017} that discusses this and introduces a method to alleviate the problem).}

\begin{figure*}[t] 
    \includegraphics[]{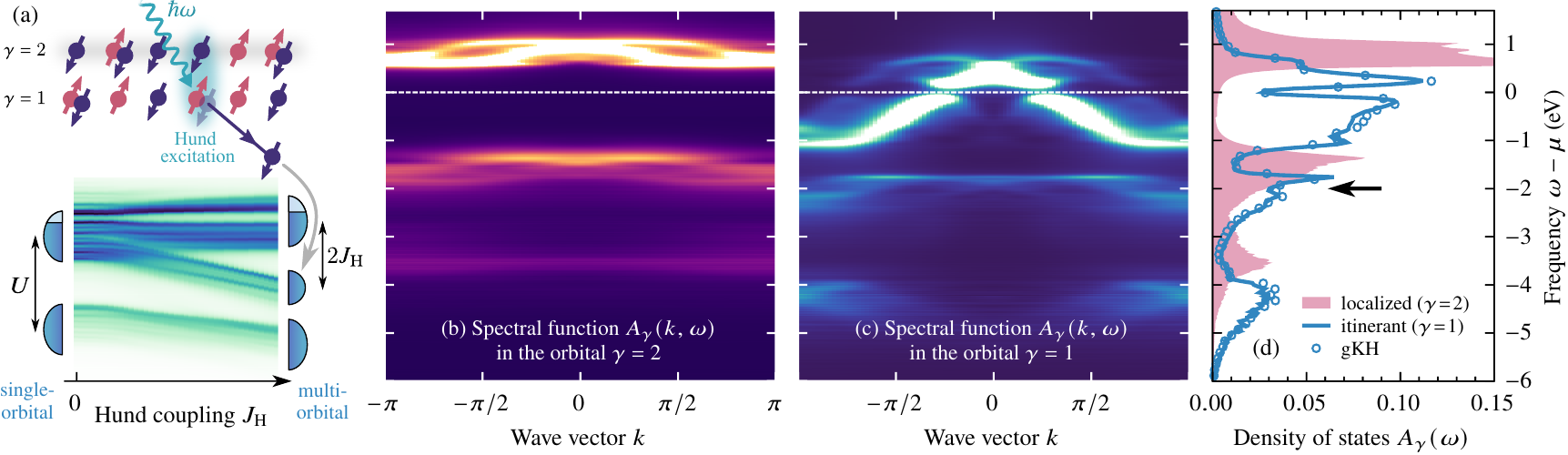}
    \caption{(a) Sketch of the Hund band accompanying the standard Hubbard bands. (b), (c) Orbital- and momentum-resolved spectral function $A_\gamma(k,\omega)$ in the two-orbital Hubbard model for $n=2.5$, $U/W=1.3$, $J_\mathrm{H}/U=0.25$, $L=48$ sites, and orbitals (b) $\gamma=2$ and (c) $\gamma=1$. The horizontal line marks the chemical potential $\mu$. (d) Orbital-resolved density-of-states $A_\gamma(\omega)$. Points depict the corresponding effective generalized Kondo-Heisenberg model (gKH); see the text for details. The arrow points at the Hund band in the itinerant orbital. Results obtained with DMRG using broadening $\eta=0.04$.}
    \label{fig1}
    \end{figure*}

In this Letter, we \ms{numerically} investigate the spectral functions of several multiorbital models. Our main result is summarized in Fig.~\ref{fig1}(a). The electronic spectrum of a single-orbital model (without the Hund coupling $J_\mathrm{H}\to0$) consists of the usual upper and lower Hubbard bands (UHB and LHB, respectively) that develop with $U$. \ms{In multiorbital systems, the finite $J_\mathrm{H}$ gives rise to additional excitations. Some of these} states can appear at energies between UHB and LHB that depend exclusively on $J_\mathrm{H}$ (i.e., are independent of $U$), paving the way to measure $J_\mathrm{H}$ directly. Since such excitations occur due to the Hund coupling and have a robust dispersion [see Fig.~\ref{fig1}(b,c) and~\cite{supp} for the full spectrum of $A(k,\omega)$], we \ms{call them} {\it Hund bands}. \jm{We recognize that the Hund bands arise whenever single-particle removal/addition processes yield a higher multiplet of the dominant valence subspace. This can occur provided: (i) the higher multiplets exist, (ii) these multiplets are allowed by the selection rules upon adding/removing a particle, and (iii) the charge fluctuations are significant. All these requirements are met for Hund's metals. Earlier work documented multiplet splittings in the Hubbard bands~\cite{Hallberg2015,Bauernfeind2017,Stadler2019}, in the fully occupied orbital~\cite{Sutter2017}, found additional ``holon-doublon'' peaks~\cite{Fernandez2018,Niu2019,Kugler2019,Komijani2019,Hallberg2020,Boidi2021}, and \ms{analyzed the energy-level structure, revealing multiplets that violate the Hund's rules~\cite{Richaud2021}}. Here, we stress that \ms{charge} excitations independent of $U$ are a \emph{generic} consequence of the multiorbital systems.}

To reach these conclusions, we use the density-matrix renormalization group method (DMRG)~\cite{White1992,Schollwock2005,White2005,Alvarez2009,Nocera2016,dmrg} and Lanczos diagonalization~\cite{Dagotto1994,Prelovsek11}. To show that our findings are generic, we \ms{study} both the two- and three-orbital Hubbard model. Furthermore, we supplement our analysis with the effective model of the OSMP - the generalized Kondo-Heisenberg Hamiltonian. Finally, we confirm our findings with DMFT calculations. Our results apply to many experiments investigating the spectral properties of multiorbital materials, particularly iron-based compounds~\cite{Watson2017,Evtushinsky2016}, ruthenates~\cite{Hotta2001,Sutter2017,Sutter2019,Gretarsson2019}, iridates~\cite{Yuan2017,Parschke2018}, and nickel oxides~\cite{Anisimov1999,Zhang2017,Li2019,Wan2021,Gu2021}.

%--------------------------------------------------------------------------------
%--------------------------------------------------------------------------------

%------------------------------------------------------------------------------------
\pagebreak
{\it Model.} 
We focus on the SU(2)-symmetric multiorbital Hubbard-Kanamori chain,
\begin{equation}
    \begin{aligned}
        H_{\mathrm{H}}=&- \sum_{\mathclap{\gamma\gamma^\prime\ell\sigma}}
        t_{\gamma\gamma^\prime}
        \left(c^{\dagger}_{\gamma\ell\sigma}c^{\phantom{\dagger}}_{\gamma^\prime\ell+1\sigma}+\mathrm{H.c.}\right)+
        \sum_{\gamma\ell} \Delta_\gamma n_{\gamma\ell}\\
        &+ U\sum_{\gamma\ell}n_{\gamma\ell\uparrow}n_{\gamma\ell\downarrow}
        +\left(U-5J_{\mathrm{H}}/2\right) \sum_{\mathclap{\gamma<\gamma^\prime\!,\ell}} n_{\gamma\ell} n_{\gamma^\prime\ell}\\
        &- 2J_{\mathrm{H}} \sum_{\mathclap{\gamma<\gamma^\prime\!,\ell}} \mathbf{S}_{\gamma\ell} \cdot \mathbf{S}_{\gamma^\prime\ell}
        +J_{\mathrm{H}} \sum_{\mathclap{\gamma<\gamma^\prime\!,\ell}} \left(P^{\dagger}_{\gamma\ell}P^{\phantom{\dagger}}_{\gamma^\prime\ell}
        +\mathrm{H.c.}\right)\!.
    \end{aligned}\vspace{2em}
    \label{hamhub}
\end{equation}
Here, $c^{\dagger}_{\gamma\ell\sigma}$ creates an electron with spin $\sigma$ at orbital $\gamma$ of site $\ell$. $t_{\gamma\gamma'}$ is the \ms{symmetric} hopping matrix in orbital space. $\Delta_\gamma$ \ms{denotes} the crystal-field splitting. $n_{\gamma\ell}=\sum_{\sigma}n_{\gamma\ell\sigma}$ represents the total density of electrons. $U$ is the standard repulsive Hubbard interaction. $J_\mathrm{H}$ is the Hund coupling between spins $\mathbf{S}_{\gamma\ell}$ at different orbitals $\gamma$. The last term $P^{\dagger}_{\gamma\ell}P^{\phantom{\dagger}}_{\gamma^{\prime}\ell}$ \ms{denotes} interorbital pair hopping, $P_{\gamma\ell}=c^{\phantom{\dagger}}_{\gamma\ell\uparrow}c^{\phantom{\dagger}}_{\gamma\ell\downarrow}$. We assume open boundary conditions, as required by DMRG. \ms{For the two-orbital model, $\gamma \in \{1,2\}$, we used (in eV): $t_{11}=-0.5$, $t_{22}=-0.15$, $t_{12}=t_{21}=0$, $\Delta_{1}=0$, $\Delta_{2}=0.8$; whereas for the three-orbital model: $\gamma \in \{0,1,2\}$, $t_{00}=t_{11}=-0.5$, $t_{22}=-0.15$, $t_{02}=t_{12}=0.1$, $t_{01}=0$, $\Delta_{0}=-0.1$, $\Delta_{1}=0$, $\Delta_{2}=0.8$.
These values were previously used to study the iron-based ladders of 123 family \cite{Herbrych2018,Herbrych2019,Rincon2014,Rincon2014-2,Sroda2021,Herbrych2020}.
The bandwidth of the two-orbital model, $W=2.1$, is used as the energy unit \cite{bandwidth}. All energy labels given throughout the text are independent of the $J_\mathrm{H}/U$ ratio.}

We also study the minimal model of the OSMP: the generalized Kondo-Heisenberg model (gKH). This model was derived~\cite{Herbrych2019,Herbrych2020,Sroda2021} to capture the static and dynamic properties of BaFe$_2$Se$_3$ iron-based ladder~\cite{Mourigal2015,Nambu2012,Wu2019}. It describes interacting itinerant electrons (with spin $\mathbf{s}_{\mathrm{i}}$) coupled via Hund coupling to the localized spins $\mathbf{S}_{\mathrm{l}}$,
\begin{equation}
    \begin{aligned}
        H_{\mathrm{K}}= &-t_{\mathrm{i}}\sum_{\ell\sigma}
        \left(c^{\dagger}_{\ell\sigma}c^{\phantom{\dagger}}_{\ell+1\sigma}+\mathrm{H.c.}\right)
        +U\sum_{\ell}n_{\ell\uparrow}n_{\ell\downarrow}\\
        &+K\sum_{\ell}\mathbf{S}_{\mathrm{l}\ell} \cdot \mathbf{S}_{\mathrm{l}\ell+1}
        -2J_{\mathrm{H}}\sum_{\ell}\mathbf{s}_{\mathrm{i}\ell} \cdot \mathbf{S}_{\mathrm{l}\ell}\,.
    \end{aligned}
\label{hamkon}
\end{equation}
\ms{For the gKH model: $t_\mathrm{i}=-0.5$, $K=4t_\mathrm{l}^2/U$, $t_\mathrm{l}=-0.15$, matching the OSMP of our two-orbital Hubbard model \cite{Herbrych2019}.}

%--------------------------------------------------------------------------------
{\it Hund \ms{bands}.} 
\ms{Let us study the orbital-resolved single-particle spectral function $A_\gamma(k,\omega)$ and the density-of-states (DOS) $A_\gamma(\omega) \propto \sum_\sigma(\langle\langle c^\dagger_{\gamma,L/2,\sigma};c^{\phantom{\dagger}}_{\gamma,L/2,\sigma} \rangle\rangle^{\mathrm{h}}_{\omega} + \langle\langle c^{\phantom{\dagger}}_{\gamma,L/2,\sigma}; c^{\dagger}_{\gamma,L/2,\sigma} \rangle\rangle^{\mathrm{e}}_{\omega})$ \cite{supp}. Here, $k$ is the momentum, $\omega$ the energy, and $\langle\langle\dots\rangle\rangle^\mathrm{h,e}_\omega$ represent the hole and electron components.}

\ms{The origin of the Hund bands can be clearly illustrated in an OSMP system.} Figure~\ref{fig1}(b)-(d) presents data for the two-orbital Hubbard model (2oH) at electron filling $n=2.5$ and interaction $U \simeq W$. Clearly, the narrow orbital $\gamma=2$ [Fig.~\ref{fig1}(b)] has a gap at the Fermi level $\mu$, while the orbital $\gamma=1$ [Fig.~\ref{fig1}(c)] is metallic with a finite DOS at $\mu$ (or a narrow pseudogap-like feature originating in the magnetic order \cite{Patel2019}). \ms{This behavior is consistent with} the OSMP~\cite{Herbrych2019}, the narrow orbital is Mott-localized with the electron density equal to~$1$. However, instead of two excitation bands (UHB and LHB), expected from the Mott physics, we observe a prominent three-peak structure [see also the DOS in Fig.~\ref{fig1}(d)]. This structure is also visible in the itinerant orbital ($\gamma=1$), Fig.~\ref{fig1}(c), with an electron density equal to $1.5$. Note that the itinerant orbital's spectrum is accurately reproduced by the effective gKH model.

%--------------------------------------------------------------------------------
\begin{figure}[!htb]
    \includegraphics[width=1.0\columnwidth]{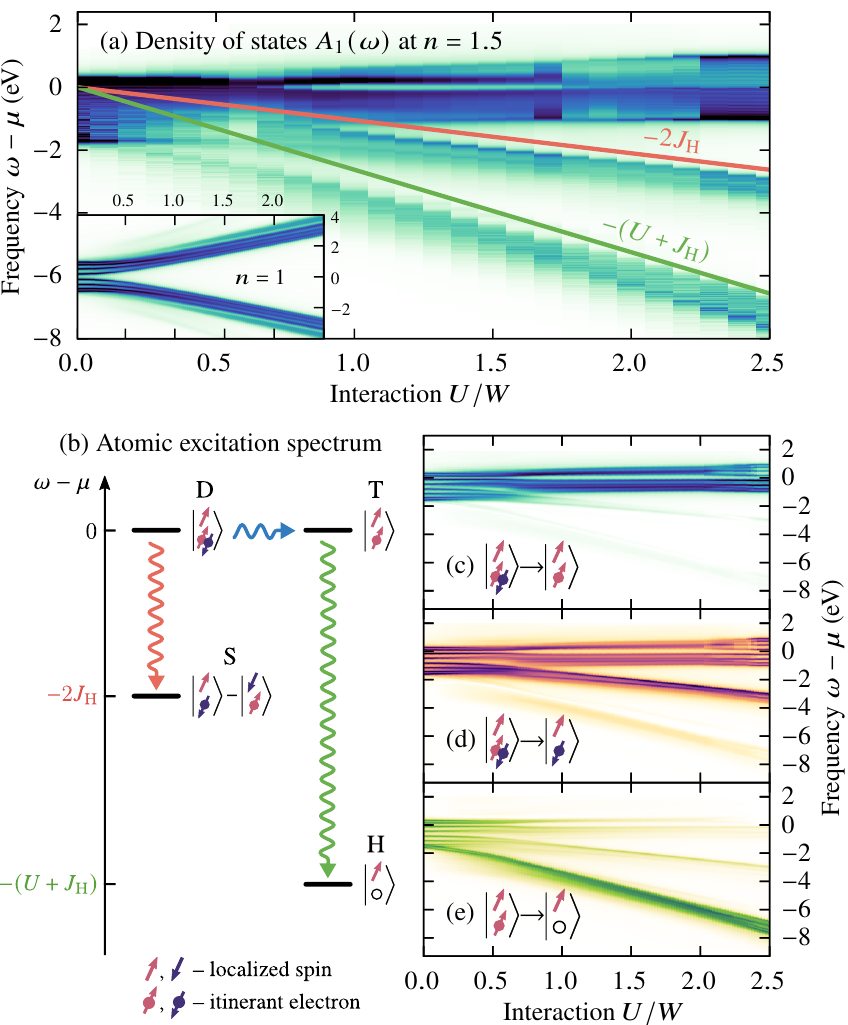}
    \caption{(a) Interaction $U$ dependence of the itinerant orbital's density-of-states (DOS) $A_1(\omega)$ obtained for the gKH model with $L=48$ sites, $J_\mathrm{H}/U=0.25$, and $n=1.5$. Results obtained with DMRG using $\eta=0.04$ broadening. The solid lines represent the atomic-limit transitions. Inset depicts the results for half electron filling $n=1$. (b) Atomic excitation spectrum. For clarity, we mark only the hole-like (electron removal) excitations and show only one spin projection. \ms{D, T, S, H labels stand for doublon, triplet, singlet, and holon, respectively.} (c)-(e) DOS $A_1(\omega)$ projected on the specific final configurations: (c) parallel spins, (d) antiparallel spins, and (e) on the holon; see the text for details. Results obtained with Lanczos diagonalization of $L=8$ lattice with broadening $\eta=0.05$.}
\label{fig2}
\end{figure}
%--------------------------------------------------------------------------------

Let us take a closer look at how the three-peak spectrum develops with the interaction $U$. Figure~\ref{fig2}(a) shows $A_1(\omega)$ for the gKH model at noninteger filling $n=1.5$. In the $U\to 0$ limit, we recover the noninteracting behavior: a single metallic band. However, already at $U/W\simeq 0.8$, i.e., close to the OSMP transition~\cite{Rincon2014,Herbrych2019}, the three-peak structure is visible in $A_1(\omega)$, and becomes clearer the larger the interaction $U$ becomes. Since the three-peak structure is most pronounced for $U\gg W$, it is instructive to examine the atomic limit \mbox{$U,J_\mathrm{H}\to \infty$} of the gKH model; see Fig.~\ref{fig2}(b). The atom realizes the noninteger filling $n=1.5$ provided the ground states (gs) of the $1$- and $2$-electron sectors are degenerate, which is achieved at $\mu=U+J_\mathrm{H}/2$. Then, the gs consists of a local interorbital triplet, denoted as $|\mathrm{T}\rangle$, which is degenerate with an itinerant doublon with localized spin, denoted as $|\mathrm{D}\rangle$. By removing an electron from the triplet, one creates a holon in the itinerant orbital ($|\mathrm{T}\rangle\to|\mathrm{H}\rangle$), with the cost of energy $U+J_\mathrm{H}$. Interestingly, from the doubly occupied state, one can remove an electron in two different ways. Depending on the spin projection of the removed electron, one can arrive at a local triplet or singlet, $|\mathrm{D}\rangle\to|\mathrm{T}\rangle$ or $|\mathrm{D}\rangle\to|\mathrm{S}\rangle$, respectively. The former is a zero-energy transition between degenerate states of the gs, while the latter costs an energy $2J_\mathrm{H}$ as it breaks the Hund's rule. In Fig.~\ref{fig2}(a), we plot the relevant energy scales of the atomic limit ($U+J_\mathrm{H}$ and $2J_\mathrm{H}$) and find good agreement with the full many-body calculations of the gKH chain. 

{\it Projections on the atomic configurations.} To make a stronger case for the atomic-limit interpretation of the three-peak spectrum, we decompose the spectral function of the full many-body calculation into individual transitions~\cite{Hallberg2020}. To this end, we use the projector ${\cal P}$ onto specific configurations of the on-site Ising basis $|\gamma=1,\gamma=2\rangle$, i.e., $\langle\langle c^\dagger_{\gamma,L/2,\sigma}; {\cal P} c^{\phantom{\dagger}}_{\gamma,L/2,\sigma} \rangle\rangle^{\mathrm{h}}_{\omega}$ \cite{supp}. For clarity, we discuss only the hole part (below~$\mu$), as the electron part can be described analogously. Upon removing an electron from the itinerant orbital, we distinguish three contributions.
(i)~In Fig.~\ref{fig2}(c), we project onto the parallel-spin configuration, ${\cal P}=|{\uparrow,\uparrow\rangle\langle\uparrow,\uparrow}|+|{\downarrow,\downarrow\rangle\langle\downarrow,\downarrow}|$. The resulting weight forms a band of excitations close to the Fermi level $\omega\simeq \mu$. This transition is responsible for the metallic properties of the lattice.
% the Hund triplets propagate in the background of doublons, which construct the UHB.
(ii)~In Fig.~\ref{fig2}(d), we instead project onto the antiparallel configuration, ${\cal P}=|{\uparrow,\downarrow\rangle\langle\uparrow,\downarrow}|+|{\downarrow,\uparrow\rangle\langle\downarrow,\uparrow}|$. We observe large weight in the middle band and some smaller weight at $\omega\simeq\mu$. The middle band represents the interorbital singlet which breaks the Hund's rule: this is the $2J_\mathrm{H}$ Hund excitation. The band at $\omega\simeq\mu$ represents the $S^z=0$ component of the triplet ($|{\uparrow,\downarrow}\rangle+|{\downarrow,\uparrow}\rangle$), costing zero energy to excite. 
(iii)~Finally, in Fig.~\ref{fig2}(e), we project onto the holon configuration, ${\cal P}=|{0,\uparrow\rangle\langle0,\uparrow}| + |{0,\downarrow\rangle\langle0,\downarrow}|$. This gives the energetically lowest band of excitations, which we recognize as the LHB, arising from triplet to holon transitions. The starting state needs to be a triplet because singlets are excluded from the gs by the Hund's rule.

%--------------------------------------------------------------------------------
{\it Noninteger vs integer filling.}
\ms{As shown above, for noninteger filling (doped system), the atomic limit is enough to explain the Hund bands.}
When the atomic gs of adjacent particle-number subspaces, say $N$ and \mbox{$N-1$}, are degenerate, there is no cost $U$ for the transition from the gs of subspace $N$ to the gs of subspace $N-1$. The excitation cost is zero; it is compensated by $\mu$ which is tuned to cause the degeneracy. However, if the $N-1$ subspace contains not only the gs but also higher multiplets, these multiplets can be accessed in the photoemission process $N\to N-1$ with just the energy $\propto J_\mathrm{H}$. Analogous reasoning applies to inverse transitions $N-1\to N$. Thus, remarkably, this results in $U$-independent Hund \ms{bands}.

\begin{figure}[tb]
    \includegraphics[width=1.0\columnwidth]{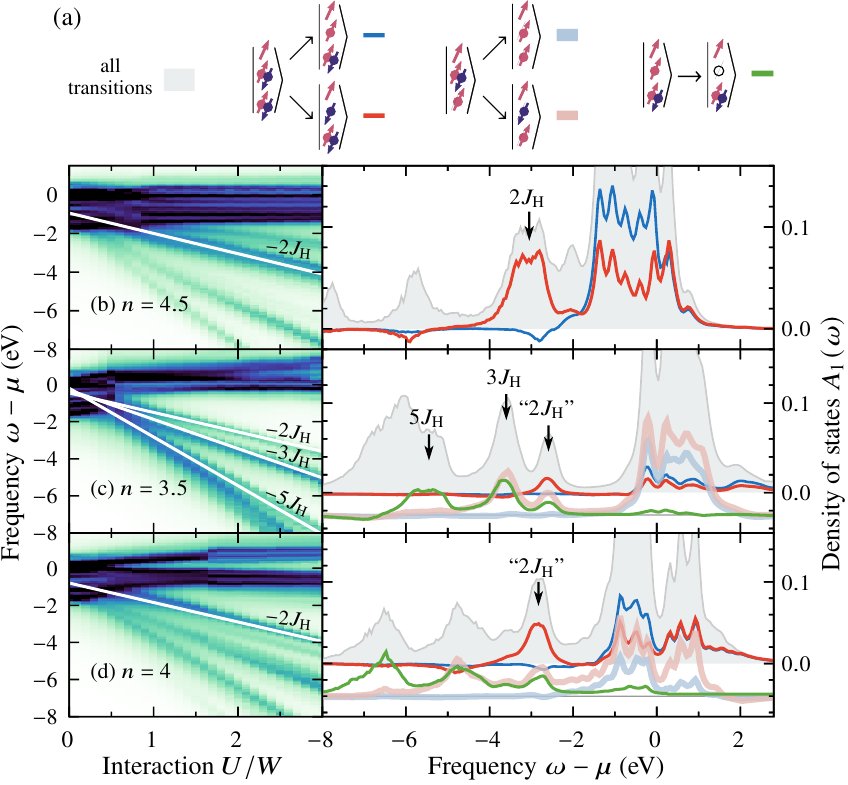}
    \caption{(a) Sketch of the transitions to final configurations that contribute to the Hund \ms{bands} of the three-orbital Hubbard model (3oH). For clarity, we present only the representative configurations (while the results are summed over several configurations of the same type). \ms{The labeling of the arrows follows Fig.~\ref{fig2}}. (b)-(d) DOS $A_1(\omega)$ of the itinerant orbital ($\gamma=1$) of 3oH with $J_\mathrm{H}/U=0.25$. Left panels depict $A_1(\omega)$ as function of the interaction $U$, while right panels show detailed spectra with projections for $U/W=2$. (b), (c), and (d) depict results for $n=4.5$, $n=3.5$, and $n=4$, respectively. The arrows on the right panels point at the Hund bands~\cite{2JH}, with ``\ldots'' denoting the peaks observable only on a lattice. The solid lines in the left column mark the $-2J_\mathrm{H}$, $-3J_\mathrm{H}$, and $-5J_\mathrm{H}$ slopes. The unlabeled peaks are Hubbard bands which have a $U$ dependence~\cite{supp}. Results obtained with DMRG on an $L=8$ lattice with broadening $\eta=0.1$.}
    \label{fig3}
\end{figure}

Consider now this behavior in a more general system, hosting more atomic configurations with different $n$. In Fig.~\ref{fig3} we present the three-orbital Hubbard model (3oH) results~\cite{3oh-osmp} for various electron fillings. For $n=4.5$, the atomic limit of our setup~\cite{supp} predicts one Hund excitation (between states with $5$ and $4$ electrons) with energy $2 J_\mathrm{H}$~\cite{2JH}, along with several $U$-dependent Hubbard excitations. We pinpoint the Hund band using the projector analysis, shown in Fig.~\ref{fig3}(b). We differentiate transitions arriving at $|{\uparrow\downarrow},{\uparrow},{\uparrow}\rangle$ and $|{\uparrow\downarrow},{\downarrow},{\uparrow}\rangle$. Similarly, for the $n=3.5$ filling, the atomic limit implies Hund bands in photoemission at $3J_\mathrm{H}$ and $5 J_\mathrm{H}$. They are shown in Fig.~\ref{fig3}(c). The $3J_\mathrm{H}$ band is a transition to a low-spin $S=1/2$ state [$\mathcal{P}$ onto $|{\uparrow},{\downarrow},{\uparrow}\rangle$; see Fig.~\ref{fig3}(a)]. The $5J_\mathrm{H}$ band originates in states of the form $|{\uparrow\downarrow},{0},{\uparrow}\rangle \pm |{0},{\uparrow\downarrow},{\uparrow}\rangle$, where ``$-$'' is degenerate with the $3J_\mathrm{H}$ excitation while ``$+$'' forms the $5J_\mathrm{H}$ peak. \ms{The latter are the holon-doublon states \cite{Fernandez2018,Niu2019,Kugler2019,Komijani2019,Hallberg2020,Boidi2021}. Their origin was discussed in~\cite{Kugler2019,Fernandez2018} but without realizing they are a particular example of the generic physics of Hund bands revealed here.} Surprisingly, the $2J_\mathrm{H}$ band persists even for $n=3.5$ (as implied by the smaller but nonvanishing weight of $|{\uparrow\downarrow},{\downarrow},{\uparrow}\rangle$), inducing a third Hund peak, absent in the atomic spectrum. The intensity of this mode decreases with $U$.

%--------------------------------------------------------------------------------
%--------------------------------------------------------------------------------

\ms{By contrast, for \emph{integer} filling $n \in \{1,2,3,\ldots\}$, the atomic limit alone does not predict the Hund bands.
The atom lacks the necessary charge fluctuations as its gs does not span adjacent particle-number subspaces.}
Thus, only the ``standard'' Hubbard bands should be observed~\cite{Medici2011,Georges2013}. However, in the lattice, the charge fluctuations are possible provided the interaction $U$ is not too large at a given filling $n$.
For half filling, the fluctuations vanish already for $U \sim W$ and the Hubbard bands are well developed  [\ms{see, e.g.,} the inset of Fig.~\ref{fig2}(a)]. Away from half filling, $U \sim W$ does not suppress the fluctuations. They are significant even at integer $n$, and vanish only at elevated $U \sim 10W$~\cite{Haule2009,Georges2013,Rincon2014-2}. \jm{Consequently, the many-body gs has significant contribution of states with neighboring local occupations, $|n-1\rangle$ and $|n+1\rangle$. Adding/removing particles in these states allows reaching the higher multiplets of the atomic ground-state subspace $|n\rangle$, and the Hund bands emerge.}

Consider the $n=4$ case, i.e., one electron above half-filling for 3oH. In the atom, the gs has only 4-electron configurations, but in the lattice we find significant on-site fluctuations to $5$- and $3$-electron states~\cite{Yin2011,Karp2020}. In Fig.~\ref{fig3}(d), we project onto the same configurations as for $n=4.5$ and again find the $2J_\mathrm{H}$ Hund band (originating in the $n=5\to4$ transitions). We should notice only half of the peak is exhausted by the projection onto $|{\uparrow\downarrow},{\downarrow},{\uparrow}\rangle$ and our results also indicate a weak $U$ dependence. \jm{We could not discern Hund bands corresponding to electron addition processes from 3-electron states: For a high-spin initial configuration $n=3, S=3/2$ the selection rules forbid reaching the low-spin $n=4, S=0$ state \cite{n3.5}.}

%--------------------------------------------------------------------------------
\begin{figure}[!t]
    \includegraphics[width=1.0\columnwidth]{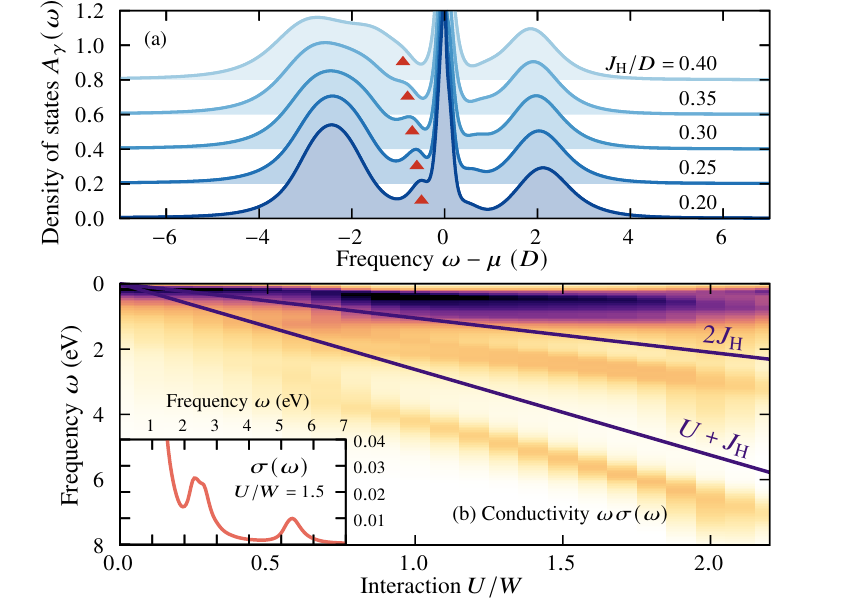}
    \caption{(a) $A_\gamma(\omega)$ calculated with the DMFT method for an orbitally degenerate three-orbital Hubbard model with semicircular DOS, integer filling $n=4$, $U/D=3.8$, and $J_\mathrm{H}/D=0.20,\dots,0.40$. The half bandwidth $D=1$ is used as the energy unit; see~\cite{supp} for details. Triangles mark $\omega=-2J_\mathrm{H}$ with a constant shift of $-0.1$~\cite{2JH}. (b)~Optical conductivity $\omega\sigma(\omega)$ vs. the interaction $U$. The lines mark the atomic-limit energy scales. Notice the $2J_\mathrm{H}$ peak appearing for $U/W>0.8$. Inset: data for $U/W=1.5$. Results obtained via DMRG for the gKH model with $n=1.5$, $J_\mathrm{H}/U=0.25$, 
$L=24$ sites, and broadening $\eta=0.1$.}
\label{fig4}
\end{figure}
%--------------------------------------------------------------------------------

%------------------------------------------------------------------------------------
{\it Conclusions.}
We showed that the charge fluctuations and finite Hund exchange present in the multiorbital Hubbard model cause the formation of unique bands of excitations.
These \ms{Hund} bands are formed by the energetically costly low-angular-momentum states (i.e., on-site configurations which break the Hund's rules) and \ms{they} do not depend on Hubbard $U$. \ms{The latter makes them} distinct from the Hubbard-band multiplet splittings. Among the Hund \ms{bands} the canonical spin-singlet mode ($\omega\simeq 2J_\mathrm{H}$) is especially prevalent. 

Our results are \ms{a generic consequence of multiorbital systems}. They originate in the \ms{existence of higher} multiplets, hence they do not depend on the presence of the orbital-selective Mott phase (see~\cite{supp} for additional discussion), nor on the system's dimensionality. To confirm this, in Fig.~\ref{fig4}(a), we present DMFT calculations in infinite dimensions. We focus on generic rather than material-specific features and consider a semicircular density of states and orbital degeneracy~\cite{supp}. The DMFT results clearly show the $2J_\mathrm{H}$ mode. \ms{In Supplemental Material \cite{supp}, we repeat the calculations for a typical $t_{2g}$ DOS and also find the Hund band.}

Our findings are relevant for ARPES, resonant inelastic x-ray scattering~\cite{Werner2021}, Raman spectroscopy~\cite{Tohyama1992,Lu2011}, nonequilibrium investigations~\cite{Strand2017,Petocchi2019,Gillmeister2020}, and reflectivity/transmission measurements~\cite{Schafgans2012,Charnukha2014,Nakajima2017,Pal2019}. Figure~\ref{fig4}(b) demonstrates the last: it presents how the optical conductivity~\cite{supp} evolves with $U$ for the gKH model at $n=1.5$. Crucially, we observe the Hund band at $\omega \simeq 2J_\mathrm{H}$. Often, such additional spectral features are attributed to the interband transitions. Here, we showed that additional modes can also originate in the Hund exchange and, consequently, can be used to estimate the value of $J_\mathrm{H}$. 

%------------------------------------------------------------------------------------
\begin{acknowledgments}
M.{\'S}. and J.H. acknowledge grant support by the Polish National Agency of Academic Exchange (NAWA) under contract PPN/PPO/2018/1/00035 and by the National Science Centre (NCN), Poland via project 2019/35/B/ST3/01207. J.M. acknowledges support by Slovenian research agency under Program No. P1-0044, J1-2458, J1-2456, J1-2463. G.A. was supported in part by the Scientific Discovery through Advanced Computing (SciDAC) program funded by the U.S. DOE, Office of Science, Advanced Scientific Computing Research and BES, Division of Materials Sciences and Engineering. E.D. was supported by the US Department of Energy, Office of Science, Basic Energy Sciences, Materials Sciences and Engineering Division. Part of the calculations has been carried out using resources provided by Wroclaw Centre for Networking and Supercomputing (\url{http://wcss.pl}).
\end{acknowledgments}
%------------------------------------------------------------------------------------

%------------------------------------------------------------------------------------
\bibliography{manuhund.bib}

\begin{thebibliography}{84}
\bibitem [{ban()}]{bandwidth}%
  \BibitemOpen
  \href@noop {} {\ }\bibinfo {note} {For the three-orbital model, the actual
  kinetic-energy span equals 2.45~eV. However, we use \mbox{$W=2.1$~eV} as the
  energy unit. This facilitates easier comparison with other
  models.}\BibitemShut {Stop}%
\bibitem [{3oh()}]{3oh-osmp}%
  \BibitemOpen
  \href@noop {} {\ }\bibinfo {note} {For our parameters, the three-orbital
  model is in the OSMP (with $\gamma=2$ localized) for $U/W \gtrsim
  0.9$.}\BibitemShut {Stop}%
\bibitem [{2JH()}]{2JH}%
  \BibitemOpen
  \href@noop {} {\ }\bibinfo {note} {We mark each Hund band by its
  $J_\mathrm{H}$ dependence. The $J_\mathrm{H}$ dependence agrees between the
  lattice and the atom very well, even though the exact values of the
  excitation energies may differ. Such a situation is clearly observed in
  Fig.~\ref{fig2}(a). In subsequent figures, we often shift the atomic energy
  scales by a constant so as to match the evolution of the lattice
  peaks.}\BibitemShut {Stop}%
\bibitem [{n3.()}]{n3.5}%
  \BibitemOpen
  \href@noop {} {\ }\bibinfo {note} {The same selection rules apply at $n=3.5$,
  where the $2J_\mathrm{H}$ and $5J_\mathrm{H}$ Hund bands of $n=4$ subspace
  are not observed in electron addition processes (i.e., above
  $\mu$).}\BibitemShut {Stop}%
\end{thebibliography}
%================================================================================
% Supplementary information
%================================================================================
\clearpage
\appendix
%\starttocentries
\setcounter{page}{1}
\setcounter{figure}{0}
\setcounter{equation}{0}
\setcounter{table}{0}
\newcommand{\rom}[1]{\uppercase\expandafter{\romannumeral #1\relax}}
\renewcommand{\refname}{Supplemental References}
\renewcommand{\figurename}{Supplemental Figure}
\renewcommand{\thefigure}{S\arabic{figure}}
\renewcommand{\theHfigure}{S\arabic{figure}} % otherwise hyperrefs point to main text
\renewcommand{\citenumfont}[1]{#1}
\renewcommand{\bibnumfmt}[1]{[#1]}
\renewcommand{\thepage}{S\arabic{page}}
\renewcommand{\theequation}{S\arabic{equation}}
\renewcommand{\theHequation}{S\arabic{equation}} % otherwise hyperrefs point to main text
\renewcommand{\thetable}{S\Roman{table}}
\renewcommand{\theHtable}{S\arabic{table}} % otherwise hyperrefs point to main text
%================================================================================
\onecolumngrid

\begin{center}
{\bf \uppercase{Supplemental Material} for}\\
\vspace{3pt}
{\bf \large \makeatletter\@title\makeatother}\\
\vspace{10pt}
by M. {\'S}roda, J. Mravlje, G. Alvarez, E. Dagotto, and J. Herbrych
\end{center}
\vspace{10pt}

\twocolumngrid

%--------------------------------------------------------------------------------
\section{Supplemental Note 1\\Definition of the single-particle spectral function}
\ms{The orbital-resolved single-particle spectral function is defined as
\begin{equation}
\resizebox{0.5\textwidth}{!}{$
    \begin{aligned}
        A_\gamma(k,\omega) &= \frac{1}{\sqrt{L}}\sum_{\ell\sigma}\mathrm{e}^{ik(\ell-c)}
        \left(\langle\langle c^\dagger_{\gamma\ell\sigma};c^{\phantom{\dagger}}_{\gamma c\sigma} \rangle\rangle^{\mathrm{h}}_{\omega}
        + \langle\langle c^{\phantom{\dagger}}_{\gamma\ell\sigma}; c^{\dagger}_{\gamma c\sigma} \rangle\rangle^{\mathrm{e}}_{\omega}
        \right),\\
        &\langle\langle A;B \rangle\rangle^{\mathrm{h,e}}_{\omega}=-\frac{1}{\pi}\operatorname{Im}\,\langle\text{gs}| A \frac{1}{\omega^{+}\pm(H- \epsilon_{\text{gs}})} B|\text{gs}\rangle\,.
    \end{aligned}
$}
\label{spectral}
\end{equation}
Here, $k$ denotes the momentum, the $+$ and $-$ signs correspond to the hole $\langle\langle \ldots \rangle\rangle^{\mathrm{h}}_{\omega}$ and electron $\langle\langle \ldots \rangle\rangle^{\mathrm{e}}_{\omega}$ components, respectively, $\omega^{+}=\omega+i\eta$ with $\omega$ being the energy, $c=L/2$, and $|\text{gs}\rangle$ is the ground state (gs) with energy $\epsilon_{\text{gs}}$. The density of states (DOS) is obtained as $A_\gamma(\omega)= (1/L) \sum_k A_\gamma(k,\omega)$.}

\ms{The origin of $c=L/2$ is the so-called central-site trick. In the Fourier transform, Eq. \eqref{spectral}, we replace the expensive double sum over two sites with a single sum over just one site. In other words, the distances are measured w.r.t. the center of the lattice. This trick is exact for periodic boundary conditions (where each site $c$ is equivalent) and introduces only minor quantitative corrections in the open boundary case.}

%--------------------------------------------------------------------------------
\section{Supplemental Note 2\\Three-orbital results for the momentum-dependence of the spectral function}
In the main text, we used the two-orbital Hubbard model (2oH) to show that the Hund excitations form robust dispersive bands. Here, we show that the same holds true for the three-orbital Hubbard model (3oH). 

\begin{figure}[!t]
    \includegraphics[width=1.0\columnwidth]{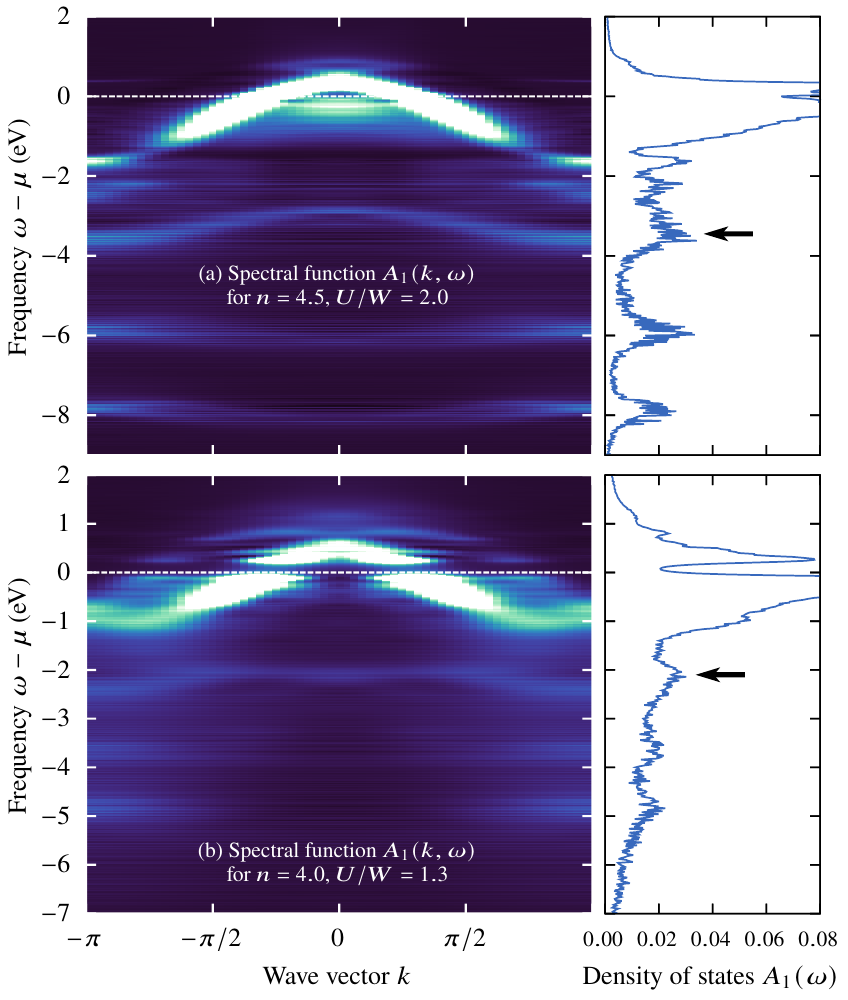}
    \caption{Momentum-resolved single-particle spectral function $A_\gamma(k,\omega)$ of the itinerant orbital $\gamma=1$ of the three-orbital Hubbard model for (a)~$n=4.5$, $U/W=2.0$ and (b)~$n=4.0$, $U/W=1.3$. The Hund coupling is $J_\mathrm{H}/U=0.25$. The right column shows the corresponding momentum-integrated spectra, i.e., the densities of states $A_\gamma(\omega)$. The arrows point at the Hund bands, as identified in the main text. Results obtained with DMRG on a $L=32$ lattice with broadening $\eta=0.04$.}
    \label{figS2}
\end{figure}

Figure~\ref{figS2} presents the momentum-resolved single-particle spectral function $A_\gamma(k,\omega)$ in the itinerant orbital $\gamma=1$ of the 3oH. Panels (a) and (b) show the fillings $n=4.5$ and $n=4.0$, respectively. Clearly, in the momentum-resolved function $A_\gamma(k,\omega)$ the spectral features are much better separated than in the density of states $A_\gamma(\omega)$. In particular, one observes that the Hund excitations (arrows) have robust dispersion and indeed form individual bands. This is particularly interesting for the integer filling $n=4.0$ [Fig.~\ref{figS2}(b)], where the Hund excitation is not a consequence of the atomic limit but is still possible due to intersite particle fluctuations.

%--------------------------------------------------------------------------------
%--------------------------------------------------------------------------------
%\pagebreak

%--------------------------------------------------------------------------------
\section{\shortstack{Supplemental Note 3\\Two- and three-orbital results for fixed $J_\mathrm{H}$}}
In this Supplemental Note, we show that the Hund bands are $U$ independent. To this end, we fix $J_\mathrm{H}=W/2$ and vary only $U$. The remaining parameters of the models are kept the same as in the main text. 

\begin{figure}[!htb]
    \includegraphics[width=1.0\columnwidth]{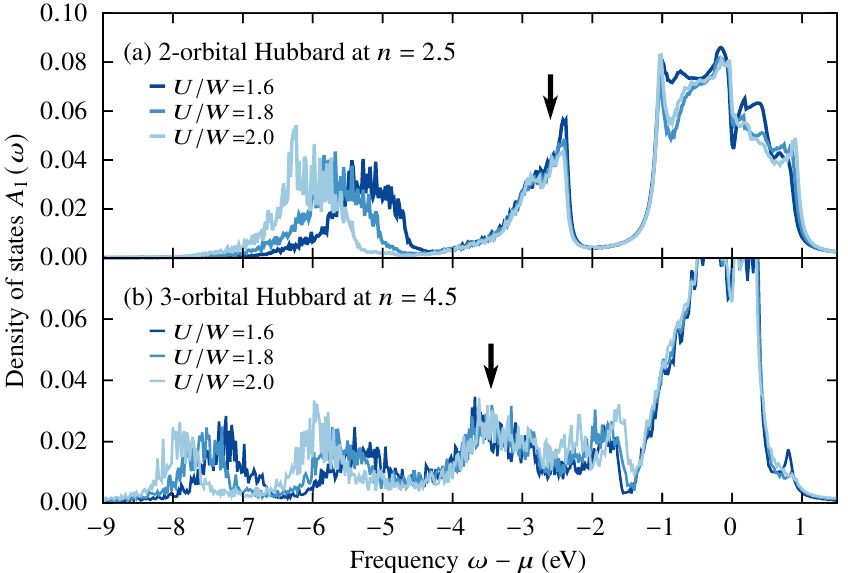}
    \caption{(a) $A_\gamma(\omega)$ of the itinerant orbital ($\gamma=1$) for the two-orbital Hubbard model ($n=2.5$, $L=48$) with fixed $J_\mathrm{H}=W/2$ and varying $U$. Clearly, the Hund band depends only on $J_\mathrm{H}$ and is $U$ independent. (b) The same as in (a) but for the three-orbital Hubbard model with $n=4.5$ and $L=32$ sites. The arrows point at the Hund bands, as identified in the main text. The unlabeled peaks are Hubbard bands which clearly have a $U$ dependence. Results obtained with DMRG using broadening $\eta=0.04$.}
    \label{figS1}
\end{figure}

Figure~\ref{figS1}(a) presents the density of states $A_1(\omega)$ of the itinerant orbital of the two-orbital Hubbard model (2oH) at $n=2.5$. Unlike the leftmost peak, which has a clear $U$ dependence, the middle peak does not shift when the interaction $U$ is varied. This is the Hund-singlet excitation with the atomic energy $2 J_\mathrm{H}$, identified in Fig.~\ref{fig1}(d) and Fig.~\ref{fig2} of the main text. Here, it can be clearly seen that its energy is fully independent of the $J_\mathrm{H}/U$ ratio.

Similar behavior is observed in the three-orbital Hubbard model (3oH). Figure~\ref{figS1}(b) shows the data for the filling $n=4.5$, where one expects the same Hund-singlet excitation as above. Among several $U$-dependent excitations (Hubbard excitations), one observes a prominent peak whose position and shape are not affected by the change in $U$. Again, this is the Hund-singlet excitation identified in Fig.~\ref{fig3}(b) of the main text.

%--------------------------------------------------------------------------------
%--------------------------------------------------------------------------------
%\pagebreak

%--------------------------------------------------------------------------------
\section{\shortstack{Supplemental Note 4\\Atomic-limit excitation energies for the three-orbital model}}
In the orbitally degenerate case of the 3oH, $t_{\gamma\gamma'} \propto \delta_{\gamma\gamma'}$, $\Delta_\gamma=0$, and for an atom, the Hubbard-Kanamori model [Eq.~\eqref{hamhub} of the main text] can be brought to a simple form~\cite{Georges2013}
\begin{equation}
    H = (U-3J_\mathrm{H}) \frac{\hat{N}(\hat{N}-1)}{2} - 2 J_\mathrm{H} \mathbf{S}^2 - \frac{J_\mathrm{H}}{2} \mathbf{L}^2 + \frac{5}{2} J_\mathrm{H} \hat{N},
    \label{hamNSL}
\end{equation}
where $\hat{N}$ is the total particle number operator, $\mathbf{S}$ the total spin operator, $\mathbf{L}$ the total orbital angular momentum operator. In this form the rotational symmetry is readily apparent. Moreover, this form allow us to classify all eigenstates (and their energies) in terms of the quantum numbers $N$, $S$, $L$. This is shown in Table~\ref{multiplet-table}, where we show both `bare' energies and also energies w.r.t.\ the atomic gs for fillings $n=4.5, 4.0, 3.5, 3.0$. To obtain the energies of the excitations, we need to subtract the energies of the gs and the target state with a sign appropriate for the electron- or hole-like single-particle excitation.

\begin{table*}[!t]
    \caption{Eigenstates and eigenvalues of the Hamiltonian \eqref{hamNSL}. For brevity, in the column `Ising basis' we show only up to two representative eigenstates from a given $N$, $S$, $L$ subspace, whereas the remaining eigenstates consist of similar Ising configurations. The shown energies are independent of the $J_\mathrm{H}/U$ ratio. In the last four columns, we show the energies w.r.t.\ the gs for fillings $n=4.5, 4.0, 3.5, 3.0$ and we restrict ourselves only to the states with $N_\mathrm{gs} \pm 1$, where $N_\mathrm{gs}$ is the number of particles in the gs. The chemical potentials corresponding to the latter fillings are $\mu_{n=4.5}=4U-7J_\mathrm{H}$, $\mu_{n=4.0}=\frac{1}{2}(7U-11J_\mathrm{H})$, $\mu_{n=3.5}=3U-4J_\mathrm{H}$, $\mu_{n=3.0}=\frac{5}{2}U-5J_\mathrm{H}$, respectively. The gs of each fixed $N$ sector is given in the first row of that sector.}
    \begin{tabular}{|c|c|c|c|cccc|}
        \hline
        \multirow{2}{*}{$N,S,L$} & \multirow{2}{*}{Degeneracy} & \multirow{2}{*}{Ising basis}                                                                                                                                                                                                                                                                                                                                                                                   & Energy                 & \multicolumn{4}{c|}{Energy $(\epsilon_{N,S,L} - \mu N) - (\epsilon_\mathrm{gs} - \mu N_\mathrm{gs})$}                                                                          \\ \cline{5-8} 
                                 &                             &                                                                                                                                                                                                                                                                                                                                                                                                                & $\epsilon_{N,S,L}$     & \multicolumn{1}{c|}{$n=4.5$}             & \multicolumn{1}{c|}{$n=4.0$}                          & \multicolumn{1}{c|}{$n=3.5$}            & $n=3.0$                           \\ \hline\hline
        $0,0,0$                  & 1                           & $|0,0,0\rangle$                                                                                                                                                                                                                                                                                                                                                                                                & $0$                    & \multicolumn{1}{c|}{}                    & \multicolumn{1}{c|}{}                                 & \multicolumn{1}{c|}{}                   &                                   \\ \hline\hline
        $1,\frac{1}{2},1$        & 6                           & $|{\uparrow},0,0\rangle$                                                                                                                                                                                                                                                                                                                                                                                       & $0$                    & \multicolumn{1}{c|}{}                    & \multicolumn{1}{c|}{}                                 & \multicolumn{1}{c|}{}                   &                                   \\ \hline\hline
        $2,1,1$                  & 9                           & \begin{tabular}[c]{@{}c@{}}$|{\uparrow},{\uparrow},0\rangle$,\\ $\frac{1}{\sqrt{2}}|{\uparrow},{\downarrow},0\rangle+\frac{1}{\sqrt{2}}|{\downarrow},{\uparrow},0\rangle$\end{tabular}                                                                                                                                                                                                                         & $U-3 J_\mathrm{H}$     & \multicolumn{1}{c|}{}                    & \multicolumn{1}{c|}{}                                 & \multicolumn{1}{c|}{$U +2J_\mathrm{H}$} & $\frac{1}{2}(U + 2J_\mathrm{H})$  \\ \hline
        $2,0,2$                  & 5                           & \begin{tabular}[c]{@{}c@{}}$\frac{1}{\sqrt{2}}|{\uparrow},{\downarrow},0\rangle-\frac{1}{\sqrt{2}}|{\downarrow},{\uparrow},0\rangle$,\\ $\frac{1}{\sqrt{6}}|{\uparrow\downarrow},0,0\rangle+\frac{1}{\sqrt{6}}|0,{\uparrow\downarrow},0\rangle-\sqrt{\frac{2}{3}}|0,0,{\uparrow\downarrow}\rangle$\end{tabular}                                                                                                & $U - J_\mathrm{H}$     & \multicolumn{1}{c|}{}                    & \multicolumn{1}{c|}{}                                 & \multicolumn{1}{c|}{$U +4J_\mathrm{H}$} & $\frac{1}{2}(U + 6J_\mathrm{H})$  \\ \hline
        $2,0,0$                  & 1                           & $\frac{1}{\sqrt{3}}|{\uparrow\downarrow},0,0\rangle+\frac{1}{\sqrt{3}}|0,{\uparrow\downarrow},0\rangle+\frac{1}{\sqrt{3}}|0,0,{\uparrow\downarrow}\rangle$                                                                                                                                                                                                                                                     & $U +  2J_\mathrm{H}$   & \multicolumn{1}{c|}{}                    & \multicolumn{1}{c|}{}                                 & \multicolumn{1}{c|}{$U +7J_\mathrm{H}$} & $\frac{1}{2}(U + 12J_\mathrm{H})$ \\ \hline\hline
        $3,\frac{3}{2},0$        & 4                           & \begin{tabular}[c]{@{}c@{}}$|{\uparrow},{\uparrow},{\uparrow}\rangle$,\\ $\frac{1}{\sqrt{3}}|{\uparrow},{\uparrow},{\downarrow}\rangle+\frac{1}{\sqrt{3}}|{\uparrow},{\downarrow},{\uparrow}\rangle+\frac{1}{\sqrt{3}}|{\downarrow},{\uparrow},{\uparrow}\rangle$\end{tabular}                                                                                                                                 & $3U - 9J_\mathrm{H}$   & \multicolumn{1}{c|}{$U -3J_\mathrm{H}$}  & \multicolumn{1}{c|}{$\frac{1}{2}(U -3J_\mathrm{H})$}  & \multicolumn{1}{c|}{$0$}                & $0$                               \\ \hline
        $3,\frac{1}{2},2$        & 10                          & \begin{tabular}[c]{@{}c@{}}$\frac{1}{\sqrt{6}}|{\uparrow},{\uparrow},{\downarrow}\rangle+\frac{1}{\sqrt{6}}|{\uparrow},{\downarrow},{\uparrow}\rangle-\sqrt{\frac{2}{3}}|{\downarrow},{\uparrow},{\uparrow}\rangle$,\\ $\frac{1}{\sqrt{2}}|{\uparrow\downarrow},0,{\uparrow}\rangle-\frac{1}{\sqrt{2}}|0,{\uparrow\downarrow},\uparrow\rangle$\end{tabular}                                                    & $3U - 6J_\mathrm{H}$   & \multicolumn{1}{c|}{$U$}                 & \multicolumn{1}{c|}{$\frac{1}{2}(U + 3J_\mathrm{H})$} & \multicolumn{1}{c|}{$3 J_\mathrm{H}$}   & $3 J_\mathrm{H}$                  \\ \hline
        $3,\frac{1}{2},1$        & 6                           & $\frac{1}{\sqrt{2}}|{\uparrow\downarrow},0,\uparrow\rangle+\frac{1}{\sqrt{2}}|0,{\uparrow\downarrow},\uparrow\rangle$                                                                                                                                                                                                                                                                                          & $3U - 4J_\mathrm{H}$   & \multicolumn{1}{c|}{$U + 2J_\mathrm{H}$} & \multicolumn{1}{c|}{$\frac{1}{2}(U + 7J_\mathrm{H})$} & \multicolumn{1}{c|}{$5 J_\mathrm{H}$}   & $5 J_\mathrm{H}$                  \\ \hline\hline
        $4,1,1$                  & 9                           & \begin{tabular}[c]{@{}c@{}}$|{\uparrow\downarrow},{\uparrow},{\uparrow}\rangle$,\\ $\frac{1}{\sqrt{2}}|{\uparrow\downarrow},{\uparrow},{\downarrow}\rangle+\frac{1}{\sqrt{2}}|{\uparrow\downarrow},{\downarrow},{\uparrow}\rangle$\end{tabular}                                                                                                                                                                & $6U - 13J_\mathrm{H}$  & \multicolumn{1}{c|}{$0$}                 & \multicolumn{1}{c|}{$0$}                              & \multicolumn{1}{c|}{$0$}                & $\frac{1}{2}(U +2J_\mathrm{H})$   \\ \hline
        $4,0,2$                  & 5                           & \begin{tabular}[c]{@{}c@{}}$\frac{1}{\sqrt{2}}|{\uparrow\downarrow},{\uparrow},{\downarrow}\rangle-\frac{1}{\sqrt{2}}|{\uparrow\downarrow},{\downarrow},{\uparrow}\rangle$,\\ $\frac{1}{\sqrt{6}}|{\uparrow\downarrow},{\uparrow\downarrow},0\rangle+\frac{1}{\sqrt{6}}|{\uparrow\downarrow},0,{\uparrow\downarrow}\rangle-\sqrt{\frac{2}{3}}|0,{\uparrow\downarrow},{\uparrow\downarrow}\rangle$\end{tabular} & $6U - 11J_\mathrm{H}$  & \multicolumn{1}{c|}{$2J_\mathrm{H}$}     & \multicolumn{1}{c|}{$2J_\mathrm{H}$}                  & \multicolumn{1}{c|}{$2J_\mathrm{H}$}    & $\frac{1}{2}(U +6J_\mathrm{H})$   \\ \hline
        $4,0,0$                  & 1                           & $\frac{1}{\sqrt{3}}|{\uparrow\downarrow},{\uparrow\downarrow},0\rangle+\frac{1}{\sqrt{3}}|{\uparrow\downarrow},0,{\uparrow\downarrow}\rangle+\frac{1}{\sqrt{3}}|0,{\uparrow\downarrow},{\uparrow\downarrow}\rangle$                                                                                                                                                                                            & $6U - 8J_\mathrm{H}$   & \multicolumn{1}{c|}{$5J_\mathrm{H}$}     & \multicolumn{1}{c|}{$5J_\mathrm{H}$}                  & \multicolumn{1}{c|}{$5J_\mathrm{H}$}    & $\frac{1}{2}(U +12J_\mathrm{H})$  \\ \hline\hline
        $5,\frac{1}{2},1$        & 6                           & $|{\uparrow\downarrow},{\uparrow\downarrow},{\uparrow}\rangle$                                                                                                                                                                                                                                                                                                                                                 & $10U - 20J_\mathrm{H}$ & \multicolumn{1}{c|}{$0$}                 & \multicolumn{1}{c|}{$\frac{1}{2}(U - 3J_\mathrm{H})$} & \multicolumn{1}{c|}{$U -3J_\mathrm{H}$} &                                   \\ \hline\hline
        $6,0,0$                  & 1                           & $|{\uparrow\downarrow},{\uparrow\downarrow},{\uparrow\downarrow}\rangle$                                                                                                                                                                                                                                                                                                                                       & $15U - 30J_\mathrm{H}$ & \multicolumn{1}{c|}{$U - 3J_\mathrm{H}$} & \multicolumn{1}{c|}{}                                 & \multicolumn{1}{c|}{}                   &                                   \\ \hline
        \end{tabular}
    \label{multiplet-table}
\end{table*}

To conduct the projector analysis, we also need to know how the eigenstates appear in the Ising-configuration basis $|\Gamma_{\gamma=0},\Gamma_{\gamma=1},\Gamma_{\gamma=2}\rangle$, where $\Gamma \in \{ |0\rangle, |{\uparrow}\rangle, |{\downarrow}\rangle, |{\uparrow\downarrow}\rangle \}$. Since the eigenstate superpositions are rather lengthy, in Table~\ref{multiplet-table} we list only up to two example eigenstates per each $N$, $S$, $L$ subspace.

Naturally, a finite crystal field ($\Delta_\gamma \neq 0$) modifies the spectrum shown in Table~\ref{multiplet-table}. The effect of the crystal field is twofold: (i) it modifies the energies and (ii) it may exclude certain configurations from the gs (more generally, it induces splittings in the spectrum). Regarding (i), if $J_\mathrm{H}, U \gg \Delta_\gamma$, the change in energies is marginal and one can still use Table~\ref{multiplet-table} to investigate the excitations. This is the case for the 3oH parameters of the itinerant orbitals, which we used in the main text. Namely, the slopes of the unlabeled excitations in Fig.~\ref{fig3} of the main text (left column) are accurately reproduced by Table~\ref{multiplet-table}. The example of effect (ii) is that in Fig.~\ref{fig3}(b), showing $n=4.5$, one cannot observe the Hund excitation with energy $5J_\mathrm{H}$ ($N,S,L=4,0,0$). To see it, the configuration $|{\uparrow\downarrow},{\uparrow},{\uparrow\downarrow}\rangle$ would need to be present in the gs, so that the photoemission from $\gamma=1$ could reach $|{\uparrow\downarrow},{0},{\uparrow\downarrow}\rangle$. However, the former configuration is excluded from the gs by $\Delta_2=0.8$ (and/or OSMP), as it has a doublon in the orbital $\gamma=2$. In contrast, such a configuration is present for the orbitally degenerate system, where the $5 J_\mathrm{H}$ excitation becomes observable (see the next Supplemental Note).

%--------------------------------------------------------------------------------
\begin{figure}[!b]
    \includegraphics[width=1.0\columnwidth]{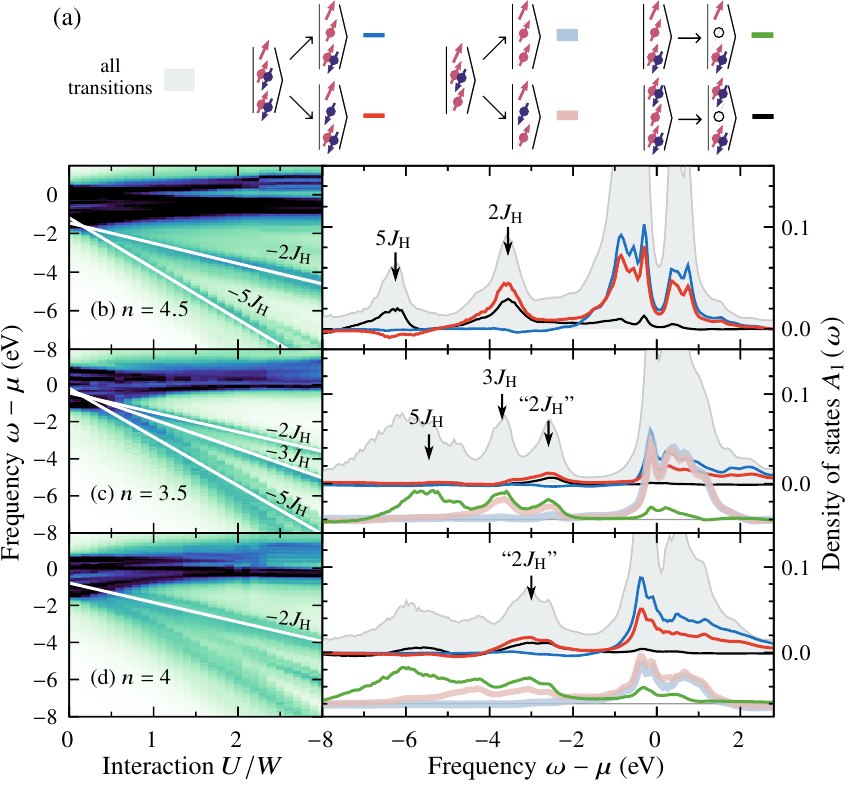}
    \caption{(a) Sketch of the transitions to final configurations that contribute to the Hund bands of the three-orbital Hubbard model (3oH). For clarity, we present only the representative configurations (while the results are summed over several configurations of the same type). (b)-(d) DOS $A_1(\omega)$ of the orbitally degenerate 3oH (see the text for details) with $J_\mathrm{H}/U=0.25$. Left panels depict $A_1(\omega)$ as a function of the interaction $U$, while right panels show detailed spectra with projections for $U/W=2$. (b), (c), and (d) depict results for $n=4.5$, $n=3.5$, and $n=4$, respectively. The arrows on the right panels point at the Hund bands~\cite{2JH}, with ``\ldots'' denoting the peaks observable only in the lattice. The solid lines in the left column mark the $-2J_\mathrm{H}$, $-3J_\mathrm{H}$, and $-5J_\mathrm{H}$ slopes. The $U$ and $J_\mathrm{H}$ dependence of the unlabeled peaks follows Table~\ref{multiplet-table}. Results obtained with DMRG on an $L=8$ lattice with broadening $\eta=0.1$.}
    \label{figS3}
\end{figure}
%--------------------------------------------------------------------------------

%--------------------------------------------------------------------------------
\section{\shortstack{Supplemental Note 5\\Hund bands in an orbitally degenerate three-orbital system}}
In this Supplemental Note, we study an orbitally degenerate 3oH (\mbox{$t_{\gamma\gamma'}=-0.5\,\delta_{\gamma\gamma'}$}, $\Delta_\gamma=0$ and other parameters the same as in the main text) via the same projector analysis as in Fig.~\ref{fig3} of the main text. Our main conclusion is that the Hund bands emerge also without the OSMP. 

In Fig.~\ref{figS3}(b), we show the results for filling $n=4.5$. According to Table~\ref{multiplet-table}, we expect two Hund excitations with energy costs $2J_\mathrm{H}$, $5J_\mathrm{H}$ and we detect both of these using appropriate projections. Surprisingly, two Hubbard excitations with energy costs $U-3J_\mathrm{H}$ and $U+2J_\mathrm{H}$, which were visible in Fig.~\ref{fig3} of the main text, are not visible here. There is also an additional weak gap forming at the chemical potential. These two features could be related to the fact that without the OSMP the system is further away from the Mott-insulating state, where the atomic limit should work best. The spectrum of the orbitally degenerate system is thus more strongly renormalized. Nonetheless, the crucial Hund excitations are clearly visible.

Figures \ref{figS3}(c),(d) display the data for fillings $n=3.5$ and $n=4.0$, respectively. Here, apart from the peaks being broader, the spectra closely resemble Fig.~\ref{fig3}(c),(d) of the main text and the same conclusions follow. Note that for $n=3.5,\, 4.0$, the energy of the $2 J_\mathrm{H}$ Hund excitation agrees with Table~\ref{multiplet-table} only if one treats the fluctuating 5-electron configuration as being part of the gs (i.e., as having energy $\epsilon_\mathrm{gs}$). For $n=4.0$ [Fig.~\ref{figS3}(d)], we also observe a small weight of the fluctuating $n=5\to4$ Hund excitation with energy $5 J_\mathrm{H}$, but for our choice of $J_\mathrm{H}/U=0.25$, it overlaps with the Hubbard excitation $\frac{1}{2}(U+7 J_\mathrm{H})$ (the slopes are very similar).

%--------------------------------------------------------------------------------
\begin{figure*}[tbh]
    \includegraphics[]{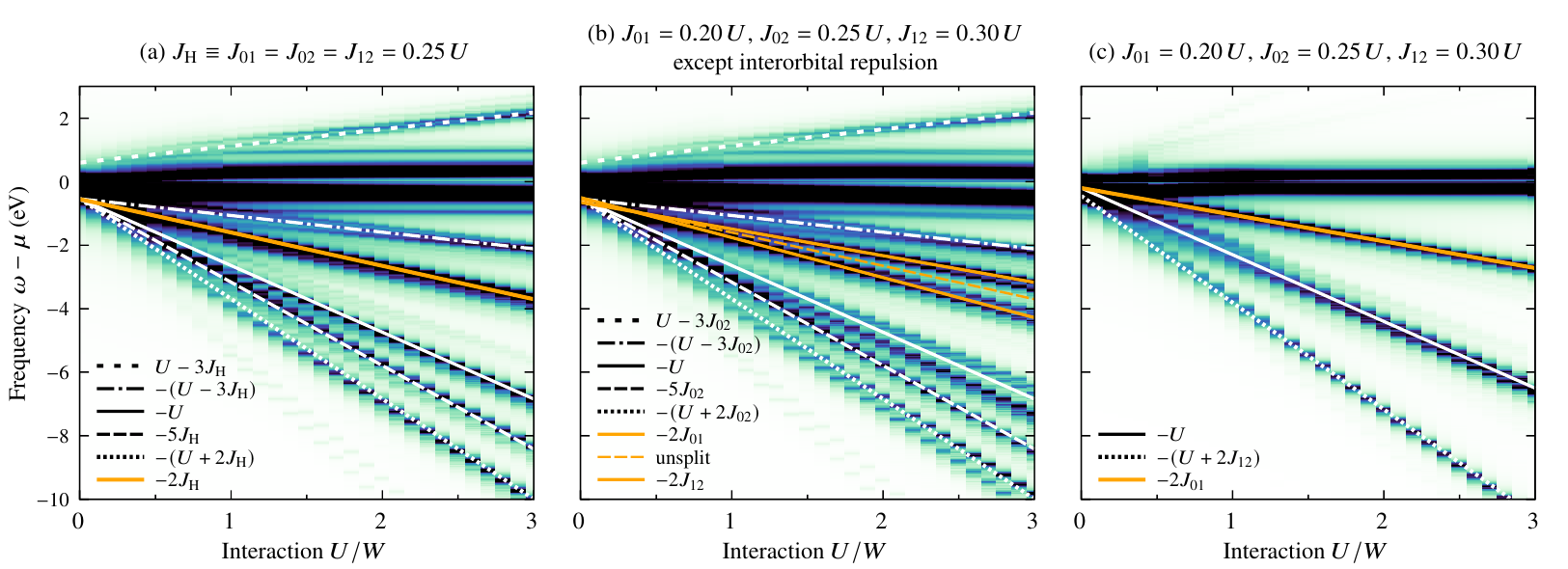}
    \caption{Density of states $A_{\gamma=1}(\omega)$ of the $\gamma=1$ orbital in the three-orbital Hubbard model. The system is orbitally degenerate and close to the atomic limit: $t_{\gamma\gamma^\prime}=0.2\delta_{\gamma\gamma^\prime}$, $\Delta_\gamma=0$, and $L=4$, with $n=4.5$. (a) Symmetric, orbitally independent Hund couplings. (b) Orbitally dependent Hund couplings, but with the interorbital repulsion kept independent. The dashed orange line corresponds to ``unsplit'' $2J_\mathrm{H}$ excitation of panel (a). (c) All interactions orbitally dependent. The lines in panels (a)-(c) correspond to excitation energies from Table~\ref{multiplet-table}. Results obtained with DMRG for broadening $\eta=0.05$.}
    \label{fig:odi}
\end{figure*}
%--------------------------------------------------------------------------------

%--------------------------------------------------------------------------------
\section{Supplemental Note 6\\Hund bands for orbital-dependent interactions}
In many materials the interactions of the Hamiltonian \eqref{hamhub} of the main text can be orbital-differentiated. To give a comprehensive answer on the fate of the Hund bands in such a case, we present additional result for the spectral functions $A_{\gamma=1}(\omega)$. We study a system where the Hund coupling depends on the orbitals involved, i.e., $J_\mathrm{H}\equiv J_{\gamma\gamma^\prime}$. In summary, two scenarios are possible: (i) the Hund band can be simply split (if the orbital occupancies remain symmetric) or (ii) the spectrum can be more strongly modified (if the occupancies become orbital-dependent).

In Fig.~\ref{fig:odi}, we show the density of states $A_{\gamma=1}(\omega)$ of the $\gamma=1$ orbital in the three-orbital Hubbard model. For the sake of clarity, we choose a simple orbital-degenerate system close to the atomic limit: $t_{\gamma\gamma^\prime}=0.2\delta_{\gamma\gamma^\prime}$, $\Delta_\gamma=0$, and $L=4$, at the $n=4.5$ filling. In Fig.~\ref{fig:odi}(a), we show the symmetric (orbital-independent) case, where all Hund couplings are the same, $J_\mathrm{H}=0.25\,U$. This case serves as a reference to the results in the main text (e.g., Fig.~\ref{fig3}) and the Supplemental Material (Fig.~\ref{figS3}). Table~\ref{multiplet-table}  accounts for all observed excitations: there is one $2J_\mathrm{H}$ line (in orange).

Next, we differentiate only the pair hopping and the direct Hund exchange: $J_\mathrm{H}\equiv J_{\gamma\gamma^\prime}$ with $\gamma=0,1,2$ denoting the orbitals. We choose $J_{01}=0.20\,U$, $J_{02}=0.25\,U$ and $J_{12}=0.30\,U$. In Fig.~\ref{fig:odi}(b), we show the scenario (i), where the $2J_\mathrm{H}$ band splits into two bands $2J_{01}$ and $2J_{12}$ (solid orange lines). The $2J_{02}$ band is not visible, because we remove an electron only from the $\gamma=1$ orbital. Thus, there are only two possible transitions giving $2J_{\gamma\gamma^\prime}$ bands, schematically shown as
\begin{eqnarray}
|{\uparrow},{\color{red}{\uparrow}{\downarrow}},{\uparrow}{\downarrow}\rangle \rightarrow |\overset{2J_{01}}{\widetilde{{\uparrow},{\color{red}{\downarrow}}}},{\uparrow}{\downarrow}\rangle\,,\nonumber\\
|{\uparrow}{\downarrow},{\color{red}{\uparrow}{\downarrow}},{\uparrow}\rangle
 \rightarrow 
|{\uparrow}{\downarrow},\overset{2J_{12}}{\widetilde{{\color{red}{\downarrow}},{\uparrow}}}\rangle\,.\nonumber
\end{eqnarray}
Besides the $2J_{01}$ and $2J_{12}$ excitations, the remaining part of the spectrum still follows Table~\ref{multiplet-table}, albeit with small splittings in its lower part. Crucially, although above we made the pair hopping and exchange interactions orbital-\emph{dependent}, we still kept the interorbital repulsion orbital-\emph{independent}. The rationale was to keep the on-site orbital occupancies intact, $\langle n_0\rangle = \langle n_1\rangle = \langle n_2\rangle =1.5$, thus keeping both on-site states $|{\uparrow},{\color{red}{\uparrow}{\downarrow}},{\uparrow}{\downarrow}\rangle$,  $|{\uparrow}{\downarrow},{\color{red}{\uparrow}{\downarrow}},{\uparrow}\rangle$ in the ground state.

In Fig.~\ref{fig:odi}(c), on the other hand, all interaction terms are orbital-dependent. Due to the now asymmetric interorbital repulsion, the system lowers its energy by unequal orbital occupancies: $\langle n_0\rangle =1$, $\langle n_1\rangle = 1.5$, and $\langle n_2\rangle =2$. Thus, the ground state keeps only the $|{\uparrow},{\color{red}{\uparrow}{\downarrow}},{\uparrow}{\downarrow}\rangle$ configuration from the two possibilities discussed above. As a result, only the $2J_{01}$ band is observable in photoemission from $\gamma=1$ orbital, whereas $2J_{12}$ is absent. Similarly, a few other bands, which were visible in Fig.~\ref{fig:odi}(a) and Fig.~\ref{fig:odi}(b), are not present in Fig.~\ref{fig:odi}(c). Their initial configurations are now excluded from the ground state. This is scenario (ii): the spectrum of a \emph{single} orbital does not include splittings and is more strongly modified due to the orbital-dependent $J_{\gamma\gamma^\prime}$. Still, if one would remove an electron from $\gamma=2$, it should be possible to observe the $2J_{02}$ band. Therefore, experimentally, when one looks at the combined spectrum of all orbitals, the splitting should be observable also in scenario (ii).

%--------------------------------------------------------------------------------
\section{\shortstack{Supplemental Note 7\\Details of the DMFT results for the semicircular DOS}}
We also considered the three-orbital problem within the DMFT method, where we used the numerical renormalization group (NRG) method~\cite{zitko09,zitko21} to solve the impurity problem. We used $\Lambda=4$, averaged over 4 different realizations of $Z$ and kept up to 4000 states in the diagonalization. We checked that the results remain consistent if these technical parameters are varied. 

\begin{figure}[t]
    \includegraphics[]{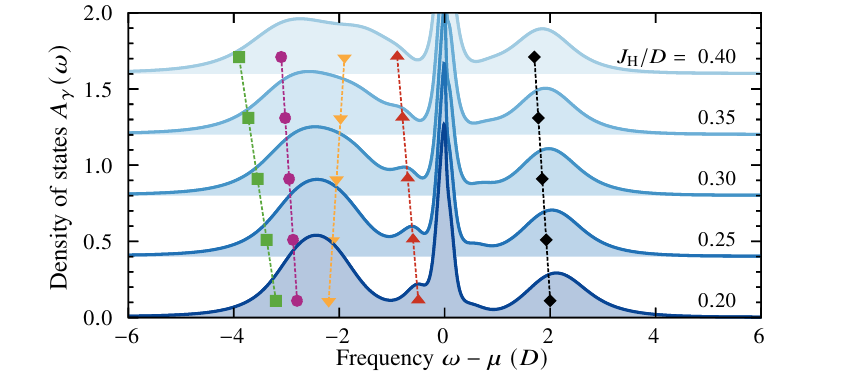}
    \caption{DMFT-NRG results for $A_\gamma(\omega)$ for an orbitally degenerate three-orbital Hubbard model with semicircular DOS of half bandwidth $D=1$ and $U/D=3.8$, $n=4$. The values of $J_\mathrm{H}$ in units of $D$ are indicated on the plot. The curves are given a vertical offset for clarity. Symbols \mbox{$\blacksquare$, $\bullet$, $\blacktriangledown$, $\blacklozenge$} mark the atomic-limit excitations \mbox{$-\frac{1}{2}(U + 7 J_\mathrm{H})-\delta$}, \mbox{$-\frac{1}{2}(U + 3 J_\mathrm{H})-\delta$}, \mbox{$-\frac{1}{2}(U -3 J_\mathrm{H})-\delta$}, and \mbox{$\frac{1}{2}(U -3 J_\mathrm{H})+\delta$}, respectively.
    Atomic-limit excitations are shifted by $\delta = D/2$; see the text for details.
    Symbol $\blacktriangle$ marks the Hund excitation $-2 J_\mathrm{H}$.
    Finally, all symbols are also shifted by $-0.1$.}
    \label{fig:dmft}
\end{figure}

In Fig.~\ref{fig4}(a) of the main text and Fig.~\ref{fig:dmft}, we show the calculated spectral functions for several $J_\mathrm{H}$. Due to the logarithmic discretization of NRG, which leads to larger broadening of the high-energy features, the Hund excitation at energy $2J_\mathrm{H}$ is best visible when it lies between the quasiparticle peak and the inner edge of the Hubbard band. This occurs for $J_\mathrm{H}/D=0.2-0.3$, where $D=1$ is the half bandwidth, used as the unit of energy. In the plot, we also indicate the values of the atomic and Hund excitations. For the atomic excitations, we additionally introduce an outward shift of $\delta=D/2$. The rationale for this shift is that the lower Hubbard band predominantly consists of the occupied states that have momenta below the Fermi surface (band energies in the window $[-D, \epsilon_\mathrm{F}]$) whereas the upper Hubbard band consists of the empty states with energies above the Fermi energy. \ms{Note that we could not discern the $5 J_\mathrm{H}$ Hund band in the DMFT data, possibly due to a combination of its small weight (it is represented by only one multiplet) and the large broadening.}

Let us also stress here that the Hund-band excitation energy grows with $J_\mathrm{H}$ and is hence distinct from the side-peak {\it within} the quasiparticle spectrum~\cite{wadati2014,stricker2014,stadler2015,horvat19,walter20} that is also characteristic of multiorbital physics. The energy of the latter feature, present in the quasiparticle peak in Hund's metals, is lower and drops with increasing $J_\mathrm{H}$.

%--------------------------------------------------------------------------------
%--------------------------------------------------------------------------------

%--------------------------------------------------------------------------------
\section{\shortstack{Supplemental Note 8\\DMFT results for a typical $t_{2g}$ DOS}}
\begin{figure}[t]
   \includegraphics[]{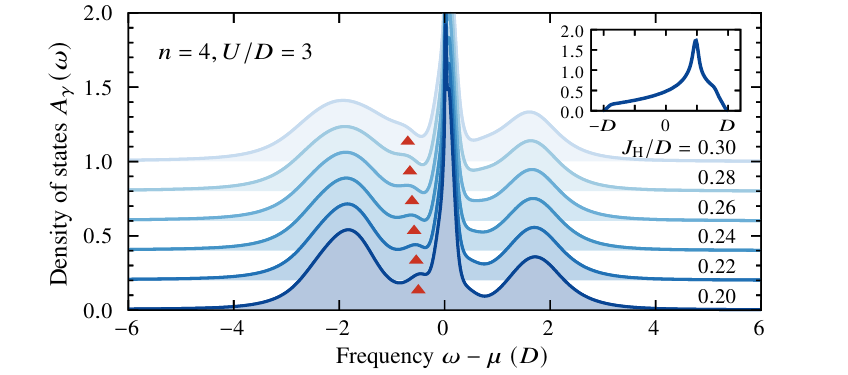}
    \caption{DMFT-NRG results for $A_\gamma(\omega)$ for an orbitally degenerate three-orbital Hubbard model with typical $t_{2g}$ DOS (inset) of half bandwidth $D=1$. The interaction and filling are $U/D=3$, $n=4$, respectively. The values of $J_\mathrm{H}$ in units of $D$ are indicated on the plot. The curves are given a vertical offset for clarity. Symbol $\blacktriangle$ marks the energy of the Hund band $-2 J_\mathrm{H} + \delta$ with the constant shift $\delta=-0.1$.}
    \label{fig:dmft2}
\end{figure}

\jm{It is of interest to investigate whether the Hund bands can be resolved also for realistic DOS. For this reason, we repeat DMFT-NRG calculations for the case of a typical $t_{2g}$ DOS (the same that was used in calculations of Ref.~\cite{Georges2013}), shown in Fig.~\ref{fig:dmft2}. One finds that the Hund bands indeed occur: the $2J_\mathrm{H}$ band is clearly present.}

\jm{Similarly as in the main text, it is advantageous to look at quite small values of $J_\mathrm{H}$ so that the peak occurs where there is no other DOS. Notice that NRG is based on logarithmic discretization, and spectra at the energies corresponding to Hubbard bands are overbroadened. If one used a solver with better resolution at higher frequencies, such as the approach of Ref.~\cite{Bauernfeind2017}, the Hund bands should be visible also when overlapping with the Hubbard bands.}

%--------------------------------------------------------------------------------
%--------------------------------------------------------------------------------

%--------------------------------------------------------------------------------
\section{\shortstack{Supplemental Note 9\\Definition of the optical conductivity}}
The calculation of the optical conductivity $\sigma(\omega)$ proved to be the most computationally demanding part of our effort. To calculate dynamical correlation functions with DMRG, we used the Krylov-space approach for correction vectors~\cite{Nocera2016}. With this method and for our system, we achieved the best results by calculating $\sigma(\omega)$ indirectly, i.e.,  from the limit of density-density correlations.

The (complex) dynamical density-density correlation function is defined as
\begin{equation}
    C_N(k,\omega^+) = -\frac{1}{\pi}\langle\mathrm{gs}| n_{\gamma k}  \frac{1}{\omega^+ - (H- \epsilon_{\text{gs}})}  n_{\gamma k} |\mathrm{gs}\rangle\,,
    \label{complex_density}
\end{equation}
where $\omega^+=\omega+i\eta$, $n_{\gamma k} = \sqrt{\frac{2}{L+1}} \sum_\ell \sin(k\ell) \left( n_{\gamma\ell} - \langle n_{\gamma\ell} \rangle \right)$ with $\langle n_{\gamma\ell} \rangle$ being the average local electron density  and $k=Z \frac{\pi}{L+1}$ ($Z=1,\ldots,L$), as appropriate for open boundary conditions. The imaginary part of Eq.~\eqref{complex_density} is the dynamical density structure factor $N(k,\omega)=\operatorname{Im}C_N$, which can be measured experimentally.

Using the continuity equation, we obtain $\sigma(\omega)$ from the long-wavelength part of the density structure factor $N(k,\omega)$~\cite{Walter1996}
\begin{equation}
    \sigma(\omega) = \lim_{k\to0} \frac{\pi}{4 \sin^2(k/2)} \omega N(k,\omega).
\end{equation}
Since in our DMRG procedure we compute both real and imaginary parts of $C_N$, we make use of both~\cite{Jeckelmann2002} and calculate the optical conductivity as 
\begin{equation}
    \sigma(\omega) = \lim_{k\to0} \frac{\pi}{4 \sin^2(k/2)} \operatorname{Im}\left( \omega^+ C_N(k,\omega^+) \right).
\end{equation}
The $k\to0$ limit is achieved by performing the calculation for the smallest possible $k$, i.e., by calculating $C_N(k=\frac{\pi}{L+1},\omega^+$).

Note that for a given system size, the above method may show larger finite-size effects than calculating $\sigma(\omega)$ directly from the current-current correlations. Still, we found it to be more computationally efficient, even though we needed a much larger bond dimension to converge $\sigma(\omega)$ than to converge the density of states $A_\gamma(\omega)$. Therefore, although our DMRG results for $\sigma(\omega)$ are qualitatively correct (they were compared to Lanczos diagonalization in small lattices), their quantitative accuracy could be further improved in future efforts.

%------------------------------------------------------------------------------------
\end{document}